\documentclass[
 aps,
 physrev,
 showkeys,
 amsmath,amssymb,
 reprint,
 unsortedaddress,
 nofootinbib,
 floatfix,
 preprintnumbers,
 noeprint,
]{revtex4-2}
\usepackage{bbm}
\usepackage{braket}
\usepackage[pdftex]{hyperref}
\usepackage{comment}
\usepackage{graphicx}
\usepackage{color}
\usepackage{orcidlink}
\usepackage{natbib}

\begin{document}

\preprint{NT@UW-25-18}

\title{Interactions of neutrino wave packets}

\author{
    Michael J.~Cervia 
    \orcidlink{0000-0002-2962-3055}
}
\email{cervia@uw.edu}
\affiliation{Department of Physics, \href{https://ror.org/00cvxb145}{University of Washington}, Seattle, Washington 98195, USA}


\begin{abstract}
The low energy effective field theory of interacting neutrinos derived from the Standard Model may be framed as a pointlike interaction and thereby modeled on a lattice of neutrino momenta. 
We identify a path to take a continuum limit of this lattice problem in the center of momentum frame. 
In this limit, the weak interaction is found to become trivial between incoming plane waves describing ultrarelativistic particles, 
unless finite neutrino wave packet sizes are taken into consideration. 
We follow up with an analytic treatment of interacting neutrino wave packets, 
demonstrating the importance 
of the wave packet size for 
characterizing neutrino-neutrino scattering in dense environments. 
\end{abstract} 

\maketitle

\section{Introduction}


Neutrino interactions are known to be weak (via $Z$ boson exchange in the Standard Model, although nonstandard interactions have been proposed and simulated extensively as well), only physically observable in sufficiently dense media (comprised of ordinary matter or a dense neutrino gas, such as the neutrino-driven wind of a supernova). 
This weak interaction, even in a limit of only forward scattering~\cite{Stanislav_P_Mikheev_1987}, permits exchange of flavor between neutrinos~\cite{PANTALEONE1992128}, allowing for collective flavor oscillations of neutrinos. 

Quantum many-body calculations, which were originally intended to emulate the collective oscillations predicted in a mean field theory and to examine the possibility of neutrino-neutrino correlations accumulating through weak flavor interaction, have been developed with increasing detail to elucidate a richer structure to collective oscillations 
(see various works over recent decades \cite{PANTALEONE1992128,
PhysRevD.48.1462, 
PhysRevD.68.013007,Alexander_Friedland_2003,LUNARDINI2005353,
PhysRevD.49.2710,PhysRevD.73.093002,PhysRevD.80.013011,
PhysRevA.65.052105,BELL200386,
PhysRevD.72.045003,
Cervia:2019nzy,Patwardhan:2019zta,Cervia:2019res,Patwardhan:2021rej,Cervia:2022pro,
Rrapaj:2019pxz,
Roggero:2021asb,Roggero:2021fyo,
Xiong:2021evk,
Lacroix:2022krq,Lacroix:2024pbb,
PhysRevD.108.123010,
Illa:2022zgu,Chernyshev:2024pqy,
Shalgar:2023ooi,Goimil-Garcia:2024wgw,
PhysRevD.110.123028,
Laraib+Sherwood2025}). 

A counterpoint to this line of work is concerned with the interaction of neutrino fields with definite momenta for periods of time much greater than flavor oscillation lengths (i.e., a toy model of ``plane waves in a box''). 
Some recent efforts have been made to reassess these correlations while modeling neutrinos of short spatial wave packet sizes, incorporated by greatly reducing the interaction time of pairs of neutrinos and evolving interactions pairwise \cite{Laraib+Sherwood2025,PhysRevD.109.103037,Shalgar:2023ooi,Goimil-Garcia:2024wgw}. 
Already, the wave packet size of neutrinos has long been argued to play a fundamental role in the mechanism of vacuum flavor oscillations \cite{PhysRevD.24.110,Akhmedov:2009rb}. 
Decoherence effects of neutrino wave packets have been considered in the context of neutrino forward scattering as well~\cite{Akhmedov:2017mcc}. 


A yet more careful and comprehensive look at how we study neutrino interactions was opened by the consideration of general momentum transfers between neutrino pairs (i.e., nonforward scattering) by Cirigliano, Sen, and Yamauchi \cite{PhysRevD.110.123028}. 
Ostensibly, the inclusion of nonforward scattering addresses a separate concern of physical validity in quantum many-body calculations from wave packet sizes, as the former calculations still consider interaction of neutrino ensembles with definite momenta (i.e., plane waves on a lattice of momenta with discrete spacing ${A}_p$ and finite volume $V_p$). 

By consideration of nonforward scattering of neutrinos with general wave packet lengths, we show that the physical importance of these effects are essentially tied to one another. 
Furthermore, we find the physical importance of neutrino-neutrino interactions to be essentially dependent upon these two effects. 

We do so first by considering a continuum limit of the lattice problem posed in nonforward scattering. 
In this limit, we come to a perhaps surprising result that the weak cross section of two \emph{ultrarelativistic} particles ($m\ll E$ for \emph{both} particles) vanishes, unless a nonvanishing wave packet size is taken into account. 

Secondly, we analytically treat the evolution of an interacting pair of neutrino wave packets of generic size $\sigma_x\sim \sigma_p^{-1}$, per the formalism of Kiers, Nussinov, and Weiss \cite{PhysRevD.53.537}. 
Within this framework, we show that both the limit of plane waves ($\sigma_p\ll p$) and the limit of narrow wave packets ($\sigma_p\gg p$) 
result in a neutrino-neutrino cross section similar to the familiar weak cross section. 
Moreover, this approach suggests that plane waves interact mostly via coherent forward scattering, while narrow wave packets could see more significant nonforward scattering. 

The rest of this article is organized as follows: 
In Sec.~\ref{sec:left}, we briefly reiterate the steps to arrive at the four-point interaction Hamiltonian for neutrinos including nonforward scattering, emphasizing the importance of approximations 
and quantization volume in doing so. 
In Sec.~\ref{sec:lattice}, we also reintroduce the lattice formalism for describing this interaction and then proceed to show new results from time evolution calculations in the center of momentum (CoM) frame. 
We reiterate the continuum limit of these lattice results via analytic treatments directly from the continuum Hamiltonian as well in Sec.~\ref{sec:cont-pw}. 
In Sec.~\ref{sec:wavepackets}, we generalize the analysis of nonforward scattering of neutrino plane waves to wave packets of general sizes. 
Before concluding, we address the separation of scales between the neutrino interaction potential and relativistic energies in Sec.~\ref{sec:K+V}. 
Finally, in Sec.~\ref{sec:concl}, we summarize this work and point to various limitations. 

\section{Effective Field Theory of the Weak Interaction Between Ultrarelativistic Particles}
\label{sec:left}

To begin our discussion, we reintroduce the full Hamiltonian for the Fermi interaction of neutrinos in general. 
We play close attention to (i) the scale $E\ll m_Z$ for low energy effective field theories in general, (ii) the ultrarelativistic scale of \emph{both} particles in the interaction $m_\nu\ll E$, and (iii) the quantization lengths introduced in reducing the full Hamiltonian to elastic or forward scattering. 

Recall the tree-level Lagrangian density for the Standard Model neutral current interaction (i.e., $Z$ boson exchange) 
\cite{PhysRevD.22.2227,PANTALEONE1992128,SIGL1993423}
\begin{align}
    \mathcal{L}_{\nu Z} = -\frac{g}{2\cos\theta_W} 
    G^{\alpha\beta} \; 
    \overline{\nu_{L,\alpha}} \gamma^\mu \nu_{L,\beta} \, Z_\mu ;
    \label{eq:nu-Z}
\end{align}
where $g$ is the weak coupling, $\theta_W$ is the Weinberg mixing angle, $\nu_L=P_L\nu = \frac{1}{2}(1-\gamma_5)\nu$ is a left-handed neutrino field in $N_f$ flavors spanned by an extra index $\alpha$ or $\beta$, $\gamma^\mu$ are the Dirac matrices, and $Z_\mu$ is the gauge field. 
Also, we have allowed for the interaction to be standard ($G=\mathbbm{1}$) or nonstandard ($G$ is Hermitian), though for simplicity and without loss of generality we consider the standard interaction for this work. 
Note that here we have taken the weak interaction still to be entirely left handed, though the presence of neutrino masses will introduce effects of a magnitude of $\mathcal{O}(m_\nu/E)$---a size of correction that we revisit in the limit of ultrarelativistic neutrino scattering, as we anticipate $m_\nu\sim10^{-2}\:\mathrm{eV}$ and $E\sim1\mathrm{-}100\:\mathrm{MeV}$ in environments that can produce copious enough neutrinos to interact coherently. 

Further, recall the low-energy effective field theory for the weak interaction, in which we expect energy transferred via neutral currents to pale in comparison to the $Z$ boson mass $m_Z \gg E/c^2$,\footnote{Given our earlier discussion, we now limit our attention to approximately left-handed neutrinos and suppress notation of their chirality as we proceed.}
\begin{align}
    \mathcal{L}(x) = 
    -\frac{G_F}{\sqrt{2}} \, 
    \overline{\nu_\alpha}(x)\gamma_\mu\nu_\beta(x) \, \overline{\nu_\beta}(x) \gamma^\mu \nu_\alpha(x)
    \label{eq:fermi4}
\end{align}
where now we have derived the pointlike, Fermi interaction $G_F = g^2 / 4\sqrt{2} \cos^2\theta_W m_Z^2 \sim 10^{-11}\:\mathrm{MeV}^{-2}$, incurring an error $\mathcal{O}(E^2/m_Z^2)$ in estimating this interaction. 
Taking the neutrino fields to be definite Weyl spinors, one can derive an effective Hamiltonian for the neutrino-neutrino interaction \cite{PhysRevD.110.123028}, 
\begin{widetext}
\begin{align}
    H_{\nu\nu} &= \frac{G_F}{\sqrt{2}} 
    \int 
    G_\mathrm{flav}^{\alpha\beta\alpha^\prime\beta^\prime}
    \, 
    G_\mathrm{mom}(\mathbf{p},\mathbf{q}, \mathbf{p^\prime}, \mathbf{q^\prime}) 
    \:
    a_{\alpha^\prime}^\dagger(\mathbf{p^\prime}) a_{\beta^\prime}^\dagger(\mathbf{q^\prime})
    a_{\alpha}(\mathbf{p}) a_{\beta}(\mathbf{q}) 
    \: (2\pi)^d \delta(\mathbf{p^\prime}+\mathbf{q^\prime}-\mathbf{p}-\mathbf{q}) \: 
    \frac{\mathrm{d}\mathbf{p}}{(2\pi)^d} \, \frac{\mathrm{d}\mathbf{q}}{(2\pi)^d} \, \frac{\mathrm{d}\mathbf{p^\prime}}{(2\pi)^d} \, \frac{\mathrm{d}\mathbf{q^\prime}}{(2\pi)^d},
    \label{eq:nunu-full}
\end{align}
\end{widetext}
where $a^\dagger$ and $a$ are dimensionful creation and annihilation operators \cite{Bjorken:1965sts} for neutrinos of definite momenta (and either flavor or mass), and we have factored couplings for momenta and flavor, 
\begin{align}
    G_\mathrm{flav}^{\alpha\beta\alpha^\prime\beta^\prime}
    &= \frac{1}{2}(\delta_{\alpha\alpha^\prime}\delta_{\beta\beta^\prime} + \delta_{\alpha\beta^\prime}\delta_{\beta\alpha^\prime}), 
    \label{eq:flav-coupling} 
    \end{align} 
    \begin{align}
    G_\mathrm{mom}(\mathbf{p},\mathbf{q},\mathbf{p^\prime},\mathbf{q^\prime}) &= f(\mathbf{p},\mathbf{q}) \, f(\mathbf{p^\prime},\mathbf{q^\prime})^*.
    \label{eq:mom-coupling}
\end{align}
Here, $f$ is a form factor dependent only on the geometry in momentum space, as derived from the appropriate inner product of Pauli spinors $\ket{\widehat{\mathbf{p}},\pm}$, 
$(\pm1)$-eigenvectors of $\widehat{\mathbf{p}}\cdot\vec{\sigma}$,
\begin{align}
    \frac{1}{\sqrt{2}} f(\mathbf{p},\mathbf{q}) &= 
    \braket{\widehat{\mathbf{p}},+|\widehat{\mathbf{q}},-} 
    \nonumber\\
    &= e^{-i\phi_{_\mathbf{p}}} \sin\left(\frac{\theta_{\mathbf{p}}}{2}\right)\cos\left(\frac{\theta_\mathbf{q}}{2}\right) 
    \nonumber \\
    &\phantom{=}- e^{-i\phi_{_\mathbf{q}}} \sin\left(\frac{\theta_{\mathbf{q}}}{2}\right)\cos\left(\frac{\theta_\mathbf{p}}{2}\right)
    \label{eq:form-fac}
\end{align}
whose magnitude takes the simple, more familiar form 
\begin{align}
    |f(\mathbf{p},\mathbf{q})|^2 = 
    1-
    \widehat{\mathbf{p}} \cdot \widehat{\mathbf{q}}.
    \label{eq:form-fac-mag}
\end{align}

Let us point out here that we arrive at this form factor $G_\mathrm{mom}$ from neglecting contributions $\mathcal{O}(m_\nu/E)$ to the spinors solving the Dirac equation, whose contractions define $f$. 
As such, we should not expect this form to apply to, e.g., neutrino-electron interactions, as even for relativistic matter we should typically have $E\sim m_e$; here, we must distinguish \emph{ultra}relativistic ($E\gg m_\nu$ neutrinos) from simply relativistic ($E\sim m$ matter). 

Now, given this general Hamiltonian for the neutrino interaction, we may restrict to two simplified cases: elastic scattering ($|\mathbf{p}|=|\mathbf{p^\prime}|$) and coherent forward scattering ($\mathbf{p}=\mathbf{p^\prime}$). 
In each case, we may obtain the simpler Hamiltonian through insertion of an appropriately normalized Dirac distribution $2\pi \ell_\mathrm{elas}^{-1} \delta(|\mathbf{p}|-|\mathbf{p^\prime}|)$ or $(2\pi)^3\ell_\mathrm{coh}^{-3}\delta(\mathbf{p}-\mathbf{p^\prime})$, respectively, with quantization lengths $\ell$ to regularize the distributions~\cite{Bjorken:1965sts,Cervia:2021qtk}. 
Thus, in the case of forward scattering, where only flavor (or equivalently momentum) is swapped, we obtain 
\begin{align}
    H_\mathrm{coh} &= 
    \frac{G_F}{\sqrt{2}V} \, G_\mathrm{flav}^{\alpha\beta\alpha^\prime\beta^\prime}
    \int 
    (1-\widehat{\mathbf{p}}\cdot\widehat{\mathbf{q}})
    \nonumber \\
    &\phantom{=\frac{G_F}{\sqrt{2}V}}
    \times 
    a_{\alpha^\prime}^\dagger(\mathbf{p}) a_{\beta^\prime}^\dagger(\mathbf{q})
    a_{\alpha}(\mathbf{p}) a_{\beta}(\mathbf{q}) \:
    \frac{\mathrm{d}\mathbf{p}}{(2\pi)^d} \, \frac{\mathrm{d}\mathbf{q}}{(2\pi)^d} ,
    \label{eq:nunu-coh} 
\end{align}
where $V$ is a spatial volume of quantization for the interacting neutrinos. 
As we reinterpret this volume in a later analysis, this volume may be thought of in terms of the size of wave packets interacting. 
Alternatively this inverse volume may be taken for interacting plane waves as a normalization related to the density of our ensemble \cite{Halzen:1984mc}. 

This part of the Hamiltonian has been studied at length in past literature, from mean-field to quantum many-body treatments. 
In the former case, coherent interactions were expected to come mainly from forward scattering of neutrinos, as the leading order term in a quantum Boltzmann equation, if the collision term is expected to be small \cite{SIGL1993423,Alexander_Friedland_2003,LUNARDINI2005353,PhysRevD.73.093002}. 
In the latter case, the estimation of some quantum correlations between interacting particles could be estimated for at least flavor swapping using this Hamiltonian \cite{Balantekin_2007}, otherwise inaccessible in a mean-field theory before collisions are incorporated. 
Originally, this term was isolated by Pantaleone \cite{PANTALEONE1992128} per an argument by Mikheev and Smirnov \cite{Stanislav_P_Mikheev_1987} applied to neutrino-matter interactions; namely, ultrarelativistic particles scattering through a thin medium would predominantly scatter forward. 
We reevaluate this intuition, restarting our analysis with the original Hamiltonian of Eq.~\eqref{eq:nunu-full}.

\section{Neutrino Interaction as a Lattice Problem}
\label{sec:lattice}


To render the problem of simulating neutrino interactions including momentum transfers computationally feasible, we discretize the allowed momentum states that neutrinos in our calculation may occupy. 
{
The momentum state of an individual neutrino $\ket{\mathbf{p}}=a^\dagger(\mathbf{p})\ket{\mathrm{vac}}$ (derived from the vacuum state $\ket{\mathrm{vac}}$) is normalized in the same way as the creation and annihilation operators, but this normalization is now controlled through the lattice spacing $A_p$. 
We express this regularization in terms of the inner product: 
$\braket{\mathbf{p}|\mathbf{q}}=(2\pi)^d\delta(\mathbf{p}-\mathbf{q})\leftarrow A_p^{-d}\delta_{\mathbf{p},\mathbf{q}}$. 
In kind, integration over real-space momentum is reduced to summation over the lattice $\mathcal{P}$ via $\int\:\mathrm{d}\mathbf{p}/(2\pi)^d\leftarrow A^d\sum_{\mathbf{p}\in\mathcal{P}}=V^{-1}\sum_{\mathbf{p}}$, in effect giving us a spatial box quantization volume $V=A_p^{-d}$. 
Notably, this lattice spacing, serving as a normalization to allowed momentum states, also plays the role of a wave packet size $\sigma_p$ limiting the spatial extent $\sigma_x\sim\sigma_p^{-1}$ of a plane wave $\ket{\mathbf{p}}$; we expound on this connection in Sec.~\ref{sec:wavepackets}. 
In this sense, one can view the continuum limit $A_p\to0$ as an expansion in our basis of allowed momentum states until they become plane waves of truly infinite (spatial) extent $V\to\infty$. 
}

In Ref.~\cite{PhysRevD.110.123028}, the lattice is squarely discretized with a spacing ${A}_p$, yet bounded within a radial cutoff $\Lambda$: $\mathcal{P} = \{\mathbf{p}={A}_p(k_x\widehat{\mathbf{p}}_x + k_y\widehat{\mathbf{p}}_y):k_x,k_y\in\mathbb{Z}, |\mathbf{p}|\leq\Lambda \}$. 
Within this model, it was found for two and for four interacting neutrinos that elastic scattering constitutes the most important part of the interaction, as the size of the neutrino-neutrino potential is much smaller than their kinetic energy. 
The elastic Hamiltonian on the lattice is thus\footnote{Of course, this Hamiltonian includes only the potential interaction, omitting the kinetic energy. For the time being, we proceed in most analyses with only this term, leaving some further considerations of kinetic energy for future work. Nevertheless, we rectify this omission ourselves and justify the applicability of our results at the end of our analysis. }
\begin{align}
    H_\mathrm{lat,el} = \frac{G_F}{\sqrt{2}V^3} G_\mathrm{flav}^{\alpha\beta\alpha^\prime\beta^\prime} \sum_{\substack{\mathbf{p},\mathbf{q},\mathbf{p^\prime},\mathbf{q^\prime}\in\mathcal{P}\\{\mathbf{p}+\mathbf{q}=\mathbf{p^\prime}+\mathbf{q^\prime}}}} 
    G_\mathrm{mom}(\mathbf{p},\mathbf{q},\mathbf{p^\prime},\mathbf{q^\prime}) \, 
    \nonumber \\
    \times 
    \delta_{|\mathbf{p}|+|\mathbf{q}|,|\mathbf{p^\prime}|+|\mathbf{q^\prime}|} a_{\alpha^\prime}^\dagger(\mathbf{p^\prime}) a_{\beta^\prime}^\dagger(\mathbf{q^\prime})
    a_\alpha(\mathbf{p}) a_\beta(\mathbf{q}) \, 
    \label{eq:nunu-elastic-lattice}
\end{align}
In principle, there are two separate limits to be taken to recover the physics of the interaction in Eq.~\eqref{eq:nunu-full}: (i) an infinite-volume limit in momentum space, $\Lambda\to\infty$ and (ii) a continuum limit, ${A}_p\to0$, each of which present computational challenges, as the Hilbert space grows with the number of allowed momenta---even with a constant number of particles occupying the lattice. 

\begin{figure}[t]
   \centering 
   \includegraphics[width=0.4\textwidth]{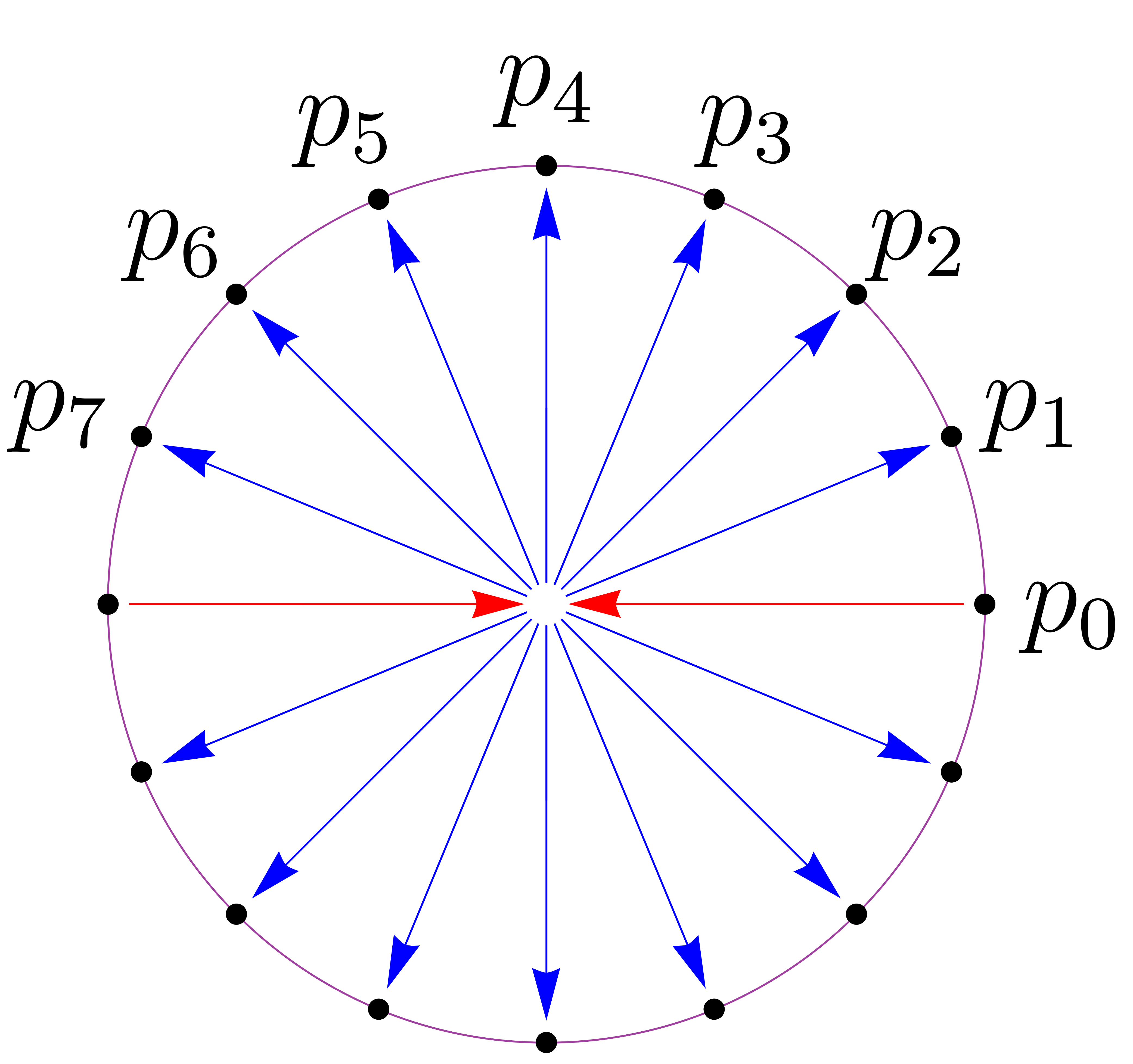}
   \caption{The space of allowed momenta resulting from the incoming state $\ket{p_0,-p_0}$ discretized to $M=8$ evenly spaced pairs. }
   \label{fig:COM_circle8}
\end{figure}

A simple scenario in which these difficulties may be addressed with relative ease involves again a pair of (standard) interacting neutrinos. 
We consider evolution on timescales long enough for interactions, which we show later to be $\sim (G_F E^3)^{-1}$ and thus much shorter than the length scale of vacuum oscillations. 
As such, we boost to the (CoM) frame for a pair of neutrinos, giving us an initial momentum state of $\ket{\mathbf{p}_0,-\mathbf{p}_0}$, and evolve with the above Hamiltonian. 

Importantly, the conservation of momentum restricts the allowed final states of this Hamiltonian to $\{\ket{\mathbf{p^\prime},-\mathbf{p^\prime}}\}$ in the CoM frame. 
Furthermore, elasticity then confines these states to antipodal pairs on the circle in momentum space $\{\ket{\mathbf{p^\prime},-\mathbf{p^\prime}}:|\mathbf{p^\prime}|=|\mathbf{p}|\}$, as depicted in Fig.~\ref{fig:COM_circle8}. 
{In effect, this restriction obviates the need for an infinite-volume limit in momentum space, allowing us to focus our efforts to a continuum limit. 
Moreover, since our volume of momentum is now a finite, fixed value $\phi$, a continuum limit may be taken via increasing the number of lattice points $M\to\infty$ such that $M /V=\phi$ is held fixed, implying $V\to\infty$ and $A_p\to0$.} 
Lastly, note that the coupling between different momenta in this frame is isotropic, $G_\mathrm{mom}\to2$, since $f(\mathbf{p},-\mathbf{p}) = \sqrt{2}$.\footnote{Here we have chosen a global phase of 1 for each Pauli spinor, $\braket{\widehat{\mathbf{p}},\pm|\widehat{\mathbf{p}},\pm}=1$, and noticed $\ket{-\widehat{\mathbf{p}},+} = \ket{+\widehat{\mathbf{p}},-}$ for definite left-handed spinors. } 
The computational difficulty of taking the necessary extrapolations from our lattice results is thus greatly simplified; we need to consider oscillations only in flavor or between different momenta pairs lying on this $|\mathbf{p}|$ circle. 
More specifically, we may consider $M$ evenly spaced pairs of points $\ket{p_k,-p_k}$ on this circle: $\mathcal{P}_\circ$ $=$ 
$\{p_k=|\mathbf{p}|(\cos\theta_k,\sin\theta_k): \theta_k=\pi k/M, \: k=0,1,\ldots,M-1\}${.} 

\begin{figure}[htb]
    \centering 
    \includegraphics[width=0.49\textwidth]{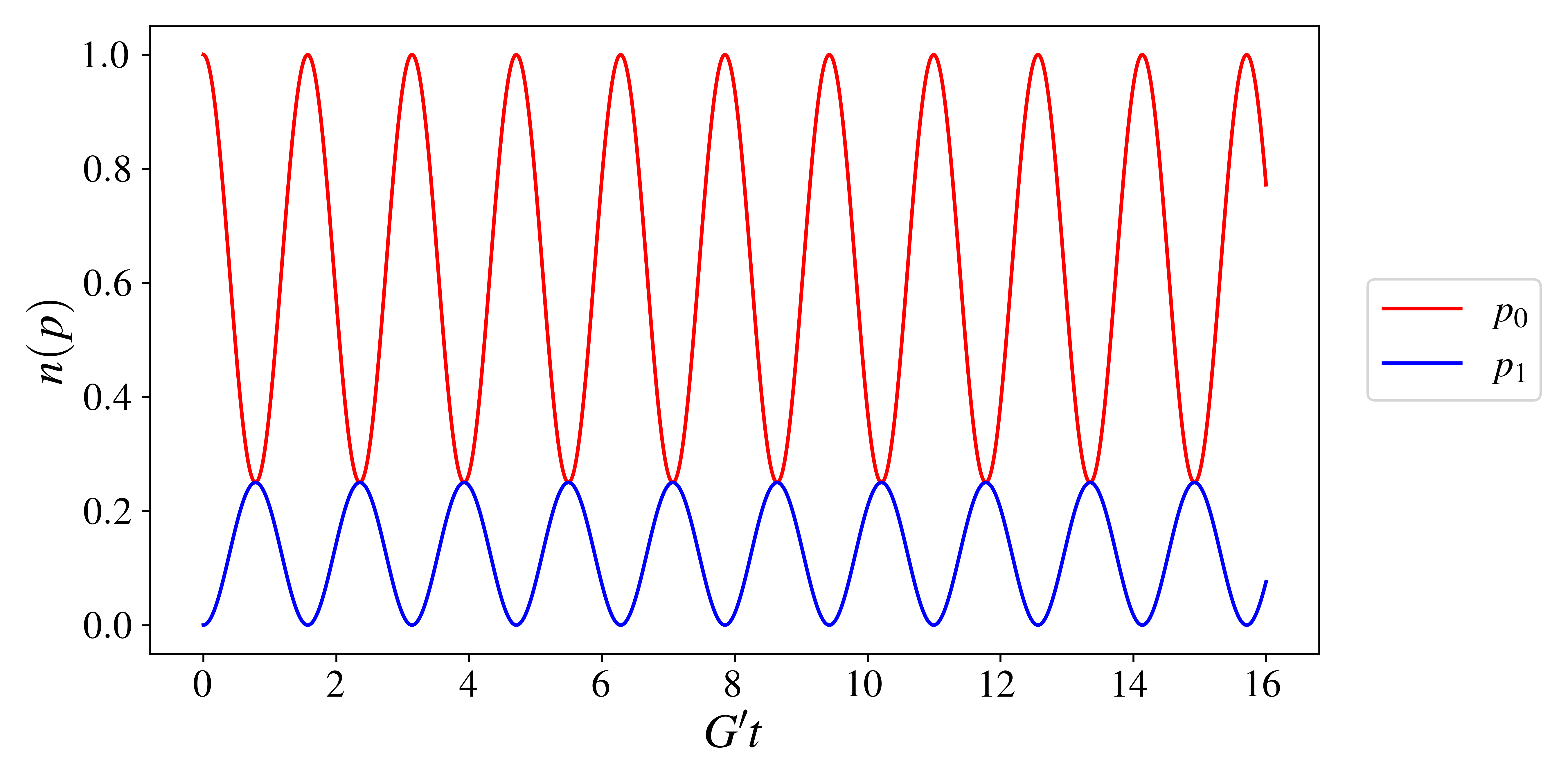}
    \\
    \includegraphics[width=0.49\textwidth]{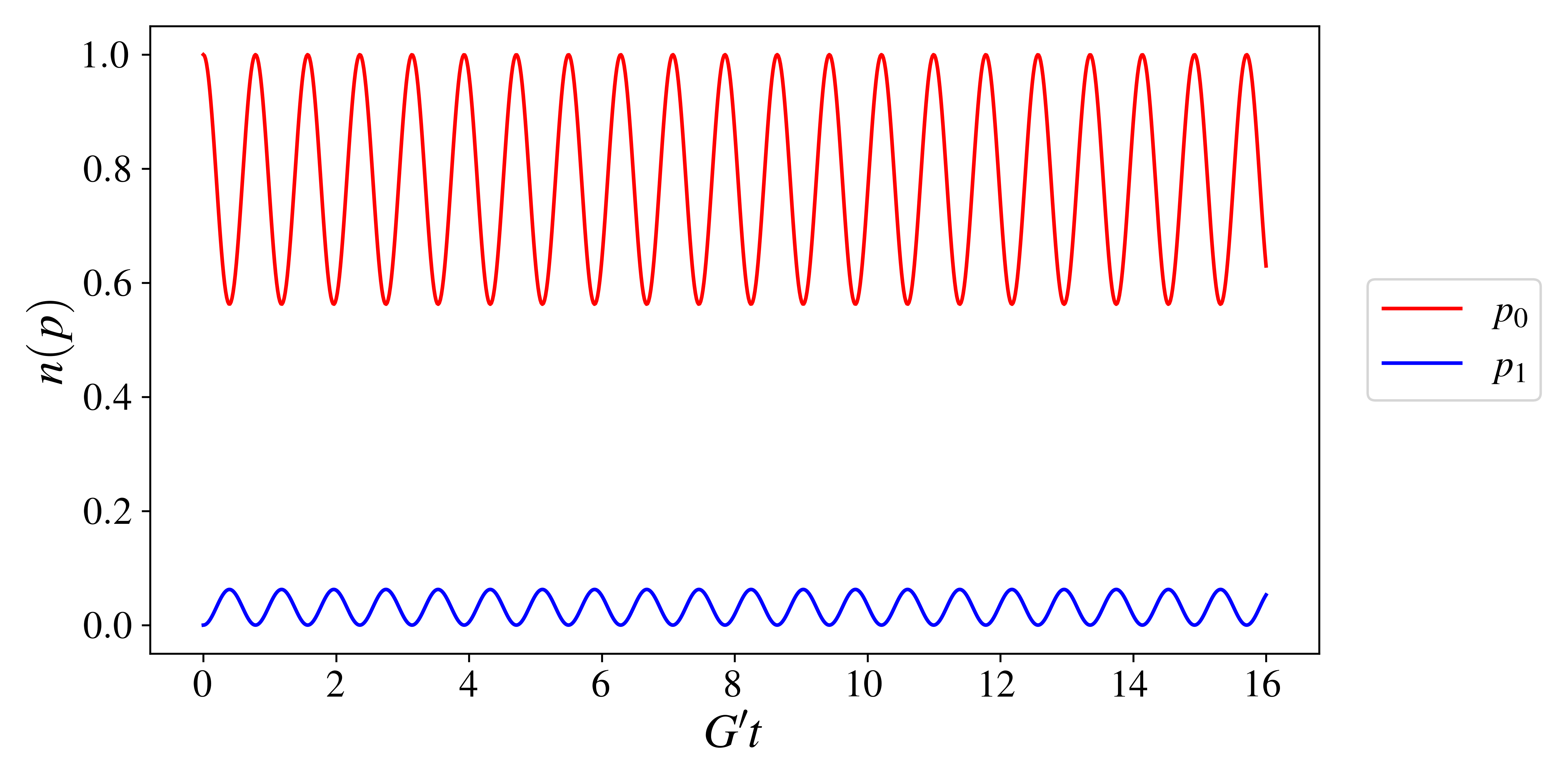}
    \caption{Oscillations over {(unitless) $G't$} time {according to Eqs.~\eqref{eq:ham-com-lat} and \eqref{eq:time-evo-lat}} in the total occupancy numbers per momentum in the lattice of the CoM frame. The incoming momentum state $\ket{p_0,-p_0}$ produces all possible outgoing states with equal likelihood, so we display just $p_1\neq p_0$ for brevity. Shown here are results from allowing for $M=4$ (top) and $M=8$ (bottom) evenly spaced pairs of outgoing momenta in simulations. {In all simulations, a dimensionless time step value of $G'\delta=0.002$ is used.} }
    \label{fig:circle_osc}
\end{figure}

In particular, let us isolate the momentum effects of our Hamiltonian in this model by evaluating the total occupancy numbers of neutrinos in each momentum state: $n(\mathbf{p})=\sum_{\alpha}n_\alpha(\mathbf{p})$, for $\mathbf{p}\in\mathcal{P}_\circ$. 
{(Since we have integrated out flavor degrees of freedom in this observable, we should expect similar results even for $N_f>2$, although comparable numerical uncertainties introduced with finite time step sizes $\delta$ may require $\delta\to (N_f/2)^{-1}\delta$.)} 
We display simulated oscillations in these occupancies for $N_f=2$ flavors over time for $M=4$ and $8$ in Fig.~\ref{fig:circle_osc}. 
It is intriguing to observe that, as the number of allowed momenta $M$ increases, the amplitude of oscillations from the initial momentum $p_0$ to $p_{i\neq0}$ gradually shrinks, while the {(dimensionless)} frequency of oscillations grows. 

\begin{figure}[t]
    \centering
    \includegraphics[width=0.48\textwidth]{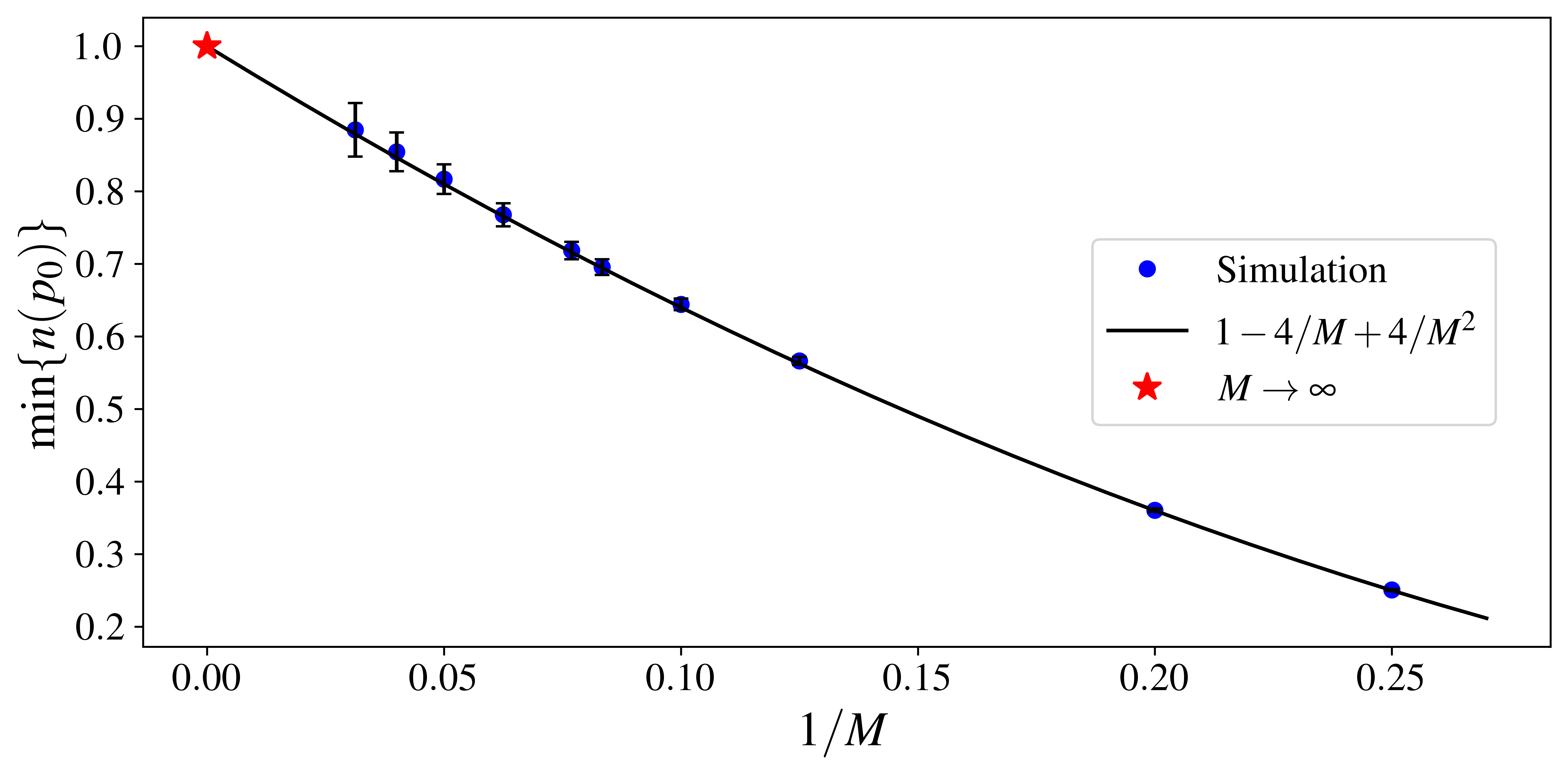}
    \\
    \includegraphics[width=0.48\textwidth]{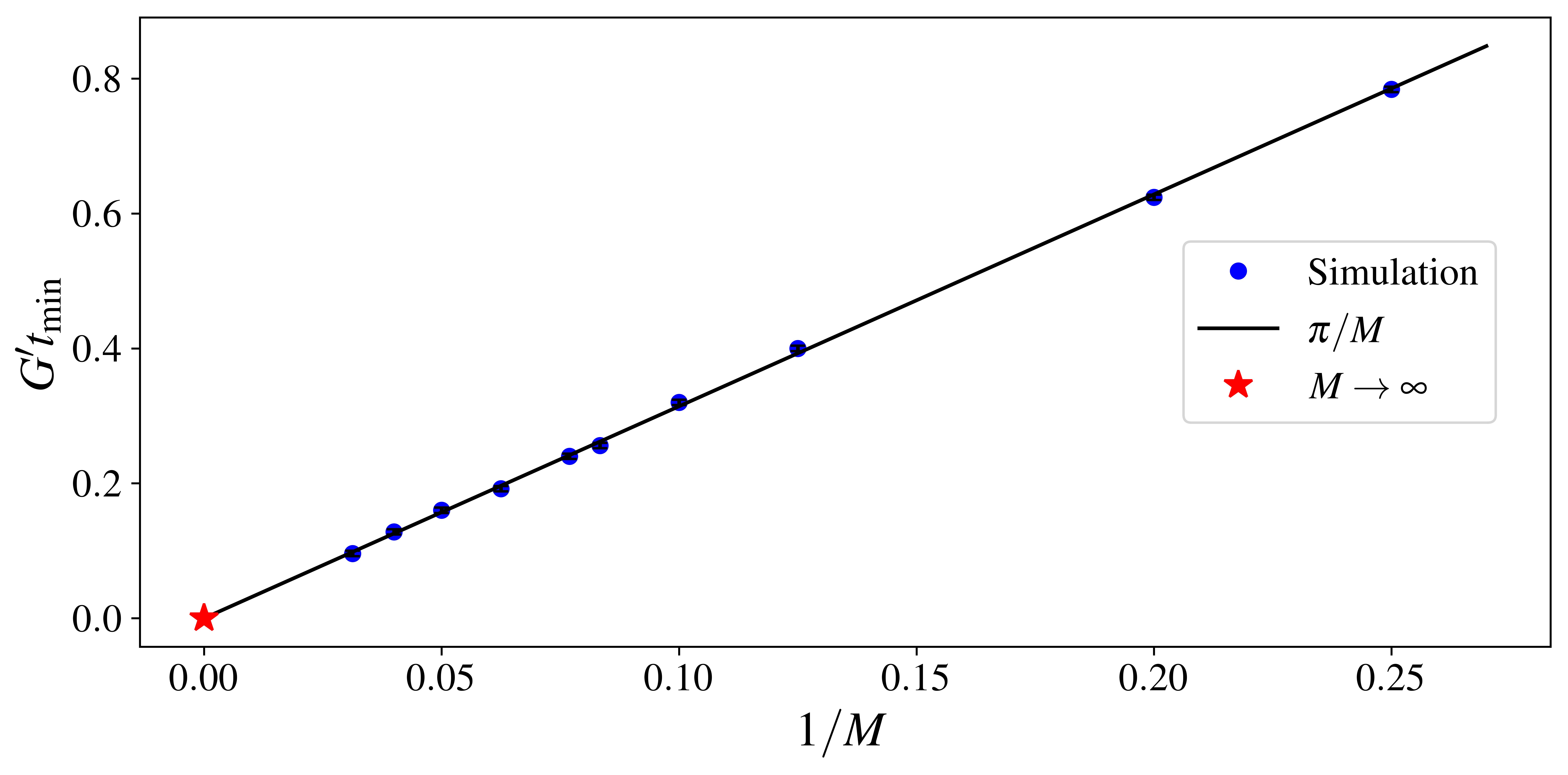}
    \caption{
    In both plots, we identify the trough in oscillations for the occupancy of the initial momentum $p_0$, $n(p_0)$, by the closest time value in numerical simulations. 
    {In all simulations, a dimensionless time step value of $G'\delta=0.002$ is used.} 
    This method produces uncertainties in measured time values and consequently in survival probabilities of simulations for each plot. 
    {Notably, numerical simulations per a program adapted from that of Ref.~\cite{PhysRevD.110.123028} are in excellent agreement with formulae derived form Eq.~\eqref{eq:time-evo-lat} (depicted as solid trend lines in each plot.)} 
    Top: we illustrate a trend in the survival probability of an incoming momentum state $p_0$ as a function of the number $M$ of allowed outgoing momentum as a result of the neutrino-neutrino interaction. 
    In fact, as we refine the discretization of our lattice in the CoM, we find this survival probability to eventually become constant (at 1) in time for $M\to\infty$. 
    Bottom: we illustrate a trend in the time required to reach the first trough in the occupancy of the incoming momentum state $p_0$ as a function of the number $M$ of allowed outgoing momentum as a result of the neutrino-neutrino interaction. We find that the oscillation frequency grows linearly with $M$, and correspondingly this required time value $t_{\min}$ trends to zero. 
    }
    \label{fig:nf-scatt_fits}
\end{figure}

We summarize these trends in $n$ and $t$ for a variety of allowed $M$ in Fig.~\ref{fig:nf-scatt_fits}. 
In fact, we can see numerically that as $M\to\infty$ there are clear trends for both the survival probability of the incoming momentum state $\ket{p_0}$ [i.e., the trough of $n(p_0)$] and the frequency of oscillations ($1/2 t_{\min}$). 
Namely, $\min\{n(p_0)\}\to1$ and ${G'}t_{\min}\to0$. 
Particularly the former trend may be counterintuitive to predict; we find that as more channels for momentum transfer are allowed, the likelihood of these momentum transfers altogether decreases. 
Making sense of this behavior and addressing this confusion is worthwhile before we proceed to more general considerations. 
In short, we see that features of these oscillations are consequences of the lattice discretization in this model. 

To understand our results in this problem, let us first summarize the simplifications of our Hamiltonian in the CoM frame as we have described above. 
First, the Hamiltonian may be factorized into a part that transfers momentum $H_\mathrm{mom}$ and a part that exchanges flavor $H_\mathrm{flav}$, comprising 
\begin{align}
    H_{\nu\nu}=H_\mathrm{flav}\otimes H_\mathrm{mom}.
    \label{eq:ham-factored}
\end{align} 
Moreover, the Hamiltonian for momentum transfer $H_\mathrm{mom}$ is isotropic in the CoM and thus takes the following form: 
\begin{align}
    H_\mathrm{mom,lat} = G'\sum_{k,l=0}^{M-1} \ket{p_k,-p_k}\!\bra{p_l,-p_l} \doteq G' \begin{pmatrix}
        1   &   \cdots  &   1   \\
        \vdots  &   \ddots  &   \vdots  \\
        1   &   \cdots  &   1
    \end{pmatrix},
    \label{eq:ham-com-lat}
\end{align}
where $G'${$= \sqrt{2} G_F/V$ (while $M/V=2\pi|\mathbf{p}|\sigma_p'^2$ is fixed for some $\sigma_p'$ characterizing any small uncertainty in the radius or width of the disk of momenta in this problem)} is an overall coupling strength.  
That is to say, our Hamiltonian ``forgets'' the directions of the initial momenta $\ket{p_0,-p_0}$ and outputs an uniform superposition of antipropagating neutrinos $\{\ket{p_k,-p_k}:k\geq0\}$.\footnote{Notably, at this stage, our argument applies to not only the earlier $d=2$ picture in our simulations, but also more general spatial dimensions $d\geq2$. } 
Mathematically, the Hamiltonian is thus proportional to a projection operator; $H_\mathrm{mom,lat}^n = (G'M)^{n-1} H_\mathrm{mom,lat}$ for $n\in\mathbb{N}$. 

It is now clear to see how this system evolves for arbitrary choices of $M$; the time evolution in momentum states thus takes the simple form 
\begin{align}
    U_\mathrm{mom,lat} = e^{-itH_\mathrm{mom,lat}} = \mathbbm{1} + (e^{-itG'M}-1)\frac{H_\mathrm{mom,lat}}{G'M}. 
    \label{eq:time-evo-lat}
\end{align}
Immediately, we may see that the (angular) frequency of oscillations is $G'M$, and crucially the evolution operator reduces to identity as $M\to\infty$. 
Observe that this trend in probability is independent of the choice of normalization in $H_\mathrm{mom}$; even if we were to adjust $G'$ as a function of $M$, the above analysis for $\min\{n(p_0)\}$ depends only on the normalized projector $H_\mathrm{mom,lat}/G'M$, whose individual matrix elements vanish as $M\to\infty$. 
In fact, we may identify the trends plotted in Fig.~\ref{fig:nf-scatt_fits} exactly using the above form for our evolution in momentum. 
The first trough of $n(p_0)$ oscillations over time will occur at $G't_{\min}=\pi/M$, and so the survival probability is $1-4/M+4/M^2$.

We should also point out that ordinary flavor swaps, resulting from forward scattering, vanish in this model as well. 
Trends similar to $n(p_0)$ can be observed for $n_e(p_0)-n_x(p_0)$ in numerical simulations of a two-flavor model, and these similarities may be expected from the factorized form of the Hamiltonian in Eq.~\eqref{eq:ham-factored}. 
The same trends are also observed when we irregularly space pairs of antipodal points on the circle of $\mathcal{P}_\circ$ (calculations not shown). 

Importantly, this limit is a consequence of how finely we resolve momentum states with our interaction on the lattice, which cannot be solved simply by binning together the probabilities of various outgoing states. 
For example, suppose we sum over all nonforward scattering outgoing states, averaged over long times; $\sum_{k>0} \overline{n}(p_k) = \max \sum_{k>0}\overline{|\braket{p_k,-p_k|e^{-itH}|p_0,-p_0}|^2} = 2(M-1)/M^2 = (1-\min\{n(p_0)\})/2 \to0$, as we can also see from the trend in the survival probability $n(p_0)$. 
We can however add \emph{coherently} the amplitudes of these outgoing nonforward scattering states for a finite result, $\sum_{k>0}\braket{p_k,-p_k|e^{-itH}|p_0,-p_0} \not\to0$ as $M\to\infty$, 
encouraging us to more deeply contemplate the meaning of our survival probabilities and amplitudes thereof calculated on the lattice. 

In order to resolve on the lattice our regular intuition that an interaction persists even in the continuum limit, we must add \emph{coherently} the wave function amplitudes across a constant fraction lattice sites in which we would like to detect a neutrino. 
That is to say, a finite resolution in momentum space is required to recover the neutrino-neutrino interacting behavior in the continuum limit. 
Let us demonstrate this interpretation first in $d=2$ and then in arbitrary dimension $d\geq2$. 

To start in $d=2$, consider a coherent sum over a finite region of the lattice in momentum space $\mathcal{P}_\angle$ $:=$ $\{|\mathbf{p}|(\cos\theta,\sin\theta):a\leq\theta\leq b\}$ containing the incoming momenta $\theta=0$ (assuming $a<0$ and $b>0)${. This sum results} 
in the overall probability 
$n(\mathcal{P}_\angle) = |\sum_{\mathbf{p'}\in\mathcal{P}_\angle} \braket{\mathbf{p^\prime},-\mathbf{p^\prime}|U_\mathrm{mom,lat}|\mathbf{p},-\mathbf{p}}|^2$ and giving a fraction of the momentum space $|\mathcal{P}_\angle|/M = |b-a|/\pi$ that is constant in $M$ with $|b-a|\sim \sigma_p/|\mathbf{p}|$.\footnote{Note that, when we choose a lattice spacing in momentum space as coarse as the wave packet size, ${A}_p\sim\sigma_p$, we recover an interpretation that a neutrino field is localized to a single site on the lattice. Otherwise, we must perform the appropriate summation over sites within a wave packet's reach. } 
Using our earlier formula in Eq.~\eqref{eq:time-evo-lat}, we find that 
\begin{align}
    \sum_{\mathbf{p'}\in\mathcal{P}_\angle\backslash\{\mathbf{p}\}} &\braket{\mathbf{p^\prime},-\mathbf{p^\prime}|U_\mathrm{mom,lat}|\mathbf{p},-\mathbf{p}} 
    \nonumber \\
    &= (e^{-itG'M}-1)\frac{|\mathcal{P}_\angle|-1}{M} \\
    \implies \sum_{\mathbf{p'}\in\mathcal{P}_\angle} 
    &\braket{\mathbf{p^\prime},-\mathbf{p^\prime}|U_\mathrm{mom,lat}|\mathbf{p},-\mathbf{p}} 
    \nonumber \\
    &= 1 + (e^{-itG'M}-1)\frac{|\mathcal{P}_\angle|}{M}. 
\end{align}
For example, if we chose the region $a=-\pi/4$, $b=\pi/4$ (a quarter of the circle) for $M\gg1$, then $|\mathcal{P}_\angle|=M/2$, showing that the coherently summed survival probability $n(\mathcal{P}_\angle)=|1+e^{-itG'M}|^2/4$ is nonvanishing in the continuum limit. 
{In principle, we can consider this volume $|\mathcal{P}_\angle|$ a wave packet size in the angle of an outgoing neutrino's momentum, over whose basis states we have summed coherently to obtain a finite result.} 
We can familiarize ourselves with this result by analogy with cross sections in scattering theory; recall that there we generally need to integrate over finite ranges of momentum for the final result to be nonvanishing as well. 

To generalize to greater dimensions, we replace the circumference of a semicircle with the solid angle of a $(d-1)$-dimensional hemisphere, $\Omega_d/2=\pi^{d/2}/\Gamma(d/2)$, and thus a solid angle per lattice site $\Omega_\mathrm{lat}=\Omega_d/2M$ over $M$ sites. 
Moreover, our coherent sum spans the points within another solid angle $\mathcal{P}_\square$ centered around $\mathbf{p}$ with size $\Omega_0$ given by $2\Omega_0/\Omega_d \sim (\sigma_p/|\mathbf{p}|)^{d-1}$,\footnote{Here, we are considering an exact CoM frame, such that there is no uncertainty in the magnitude of $|\mathbf{p}|$; if we are to introduce uncertainty here as well with the same size $\sigma_p$, then we consider a thin volume instead of a definite solid angle and effectively replace our fraction with $2\Omega_0/\Omega_d\sim(\sigma_p/|\mathbf{p}|)^{d}$. The analysis that follows remains unchanged. } so that $\Omega_0/\Omega_\mathrm{lat}$ grows linearly in $M\gg1$. 
Then, we find our result to generalize to 
\begin{align}
    \sum_{\mathbf{p'}\in\mathcal{P}_\square} \braket{\mathbf{p^\prime},-\mathbf{p^\prime}|U_\mathrm{mom,lat}|\mathbf{p},-\mathbf{p}} &= 1 + (e^{-itG'M}-1)\frac{2\Omega_0}{\Omega_d}, 
\end{align}
again yielding a finite survival probability over the wave packet $n(\mathcal{P}_\square)$ {even in the continuum limit of $\Omega_\mathrm{lat}\to0$ (or, equivalently, $M\to\infty$)}. 
In general, we find that it is vital to consider the finite width of a wave function for outgoing particles in the neutrino-neutrino interaction; only in this way do we recover a nontrivial physical prediction from this Standard Model interaction. 
{We have seen thus far on the lattice that a coherent sum over states encompassed by an outgoing wave packet permits a finite prediction for the interaction. 
Calculations of only plane waves result in a trivial interaction in a continuum limit, as their width in momentum is reduced to zero in this case. 
We may find similar physical behaviors for the time evolution of approximate plane waves, if we repeat our calculation starting from the continuum Hamiltonian of Eq.~\eqref{eq:nunu-full}, as we see in the following section. 
}

\section{Plane Wave Neutrinos Interacting in the Continuum}
\label{sec:cont-pw}

The calculations performed above begin with a lattice model approximating our original Hamiltonian of Eq.~\eqref{eq:nunu-full} and attempt to restore the continuum limit for a simple CoM scattering problem. 
Let us now revisit the original Hamiltonian and 
briefly reiterate the above argument, without the use of a lattice, to illustrate the need for more careful consideration of quantization volume (i.e., wave packet size) when treating this problem. 

Again, we consider the factored Hamiltonian for momentum transfer between a pair of neutrinos
\begin{align}
    H_\mathrm{mom} = \frac{G_F}{\sqrt{2}} \int 
    f(\mathbf{p},\mathbf{q})f(\mathbf{p^\prime},\mathbf{q^\prime})^* \: 
    a^\dagger(\mathbf{p^\prime}) a^\dagger(\mathbf{q^\prime}) a(\mathbf{p}) a(\mathbf{q}) \: 
    \nonumber \\
    \times
    (2\pi)^d \delta(\mathbf{p}+\mathbf{q}-\mathbf{p^\prime}-\mathbf{q^\prime}) \,  \frac{\mathbf{d}\mathbf{p}}{(2\pi)^d} \frac{\mathbf{d}\mathbf{q}}{(2\pi)^d} \frac{\mathbf{d}\mathbf{p^\prime}}{(2\pi)^d} \frac{\mathbf{d}\mathbf{q^\prime}}{(2\pi)^d}
    \label{eq:hmom}
\end{align}
acting on CoM states $\ket{\mathbf{p}_0,-\mathbf{p}_0}$,\footnote{These quantum states of definite momenta are defined such that $\braket{\mathbf{p}|\mathbf{p^\prime}} = (2\pi)^d\,\delta(\mathbf{p}-\mathbf{p^\prime})$. 
In principle, they also require some normalization, given usually by a quantization volume in position space, $V$ (see, e.g., Ref.~\cite{Halzen:1984mc}). 
This volume becomes all the more essential to avoid ill-defined squares of Dirac distributions when calculating inner products of states of equal and opposite momenta, $\braket{\mathbf{p},-\mathbf{p}|\mathbf{p^\prime},-\mathbf{p^\prime}} \sim V^{d} (2\pi)^d \delta(\mathbf{p}-\mathbf{p^\prime})$. }
\begin{align}
    H_\mathrm{mom} \ket{\mathbf{p}_0,-\mathbf{p}_0} 
    &= \sqrt{2} G_F \int 
    \ket{\mathbf{p^\prime},-\mathbf{p^\prime}} \: 
    \frac{\mathbf{d}\mathbf{p^\prime}}{(2\pi)^d}. 
    \label{eq:hmom-com-state}
\end{align}
As in the lattice calculation, we find $H_\mathrm{mom}$ ``forgets'' the initial direction of the incoming neutrinos in the CoM frame. 
Consequently, we may observe a similar idempotence property in the continuous space of CoM states, 
\begin{align}
    H_\mathrm{mom}^n\ket{\mathbf{p}_0,-\mathbf{p}_0} = 
    (\sqrt{2}G_F\phi_0)^{n-1}
    H_\mathrm{mom}\ket{\mathbf{p}_0,-\mathbf{p}_0},
\end{align}
where $\phi_0 := \int \mathrm{d}\mathbf{p}/(2\pi)^3$ is a volume of the outgoing momenta space, which must be renormalized (e.g., via a cutoff $\Lambda$) to obtain a finite result. 
The evolution of a state in the CoM frame again becomes easy to compute, 
\begin{align}
    U_\mathrm{mom}&\ket{\mathbf{p}_0,-\mathbf{p}_0} 
    \nonumber\\
    &= \left[\mathbbm{1} + (e^{-it\sqrt{2}G_F\phi_0}-1)\frac{H_\mathrm{mom}}{\sqrt{2}G_F\phi_0}\right]\ket{\mathbf{p}_0,-\mathbf{p}_0}. 
\end{align}
Thus, we recover the same trivial behavior in the infinite-volume limit $\phi_0\to\infty$ as we had seen in the continuum limit of the earlier lattice calculation. 
Conversely, we can begin to resolve this paradoxical behavior and recover nontrivial interaction through the more careful treatment of the momentum volume. 


Indeed, for the general initial state of plane waves $\ket{\mathbf{p}_0,\mathbf{q}_0}$ with total momentum $\mathbf{p}_\mathrm{tot}=\mathbf{p}_0+\mathbf{q}_0$, we may generalize the result of Eq.~\eqref{eq:hmom-com-state}, 
\begin{align}
    H_\mathrm{mom} \ket{\mathbf{p}_0,\mathbf{q}_0} 
    = \frac{G_F}{\sqrt{2}} &f(\mathbf{p}_0,\mathbf{q}_0) 
    \int 
    f(\mathbf{p^\prime},\mathbf{p}_\mathrm{tot}-\mathbf{p^\prime})^* 
    \nonumber\\
    &\times
    \ket{\mathbf{p^\prime},\mathbf{p}_\mathrm{tot}-\mathbf{p^\prime}} \: 
    \frac{\mathbf{d}\mathbf{p^\prime}}{(2\pi)^d}, 
    \label{eq:hmom-state}
\end{align}
implying 
\begin{align}
    H_\mathrm{mom}^n \ket{\mathbf{p}_0,\mathbf{q}_0} 
    &= \left[\sqrt{2}G_F\phi(\mathbf{p}_\mathrm{tot})\right]^{n-1}
    H_\mathrm{mom} \ket{\mathbf{p}_0,\mathbf{q}_0}, 
    \label{eq:hmom-pow-state} \\
    \mathrm{where}\quad \phi(\mathbf{p}_\mathrm{tot}) :\!&= \frac{1}{2}\int |f(\mathbf{p^\prime},\mathbf{p}_\mathrm{tot}-\mathbf{p^\prime})|^2 \: \frac{\mathrm{d}\mathbf{p^\prime}}{(2\pi)^d},  
    \label{eq:mom-vol}
\end{align}
and therefore
\begin{align}
    \label{eq:cont-evo}
    U_\mathrm{mom} \ket{\mathbf{p}_0,\mathbf{q}_0} = 
    \left[\mathbbm{1} + (e^{-it\sqrt{2}G_F\phi}-1)\frac{H_\mathrm{mom}}{\sqrt{2}G_F\phi}\right]\ket{\mathbf{p}_0,\mathbf{q}_0}. 
\end{align}
We now have a similar interesting result in terms of the volume of the space of outgoing momentum $\phi(\mathbf{p}_\mathrm{tot})$, where $|\mathbf{p}_\mathrm{tot}|$ is the scale for this volume independent of the renormalization of this integral, now keeping our result finite. 
{Here, $\sqrt{2}G_F\phi$ plays the same dual role as did $G'M$ in the previous lattice calculation of both the phase for oscillations in momentum and the normalization of the Hamiltonian term of the evolution formula [c.f., Eq.~\eqref{eq:time-evo-lat}].} 
As we show in the Appendix, the volume can indeed be renormalized and found to be $|\mathbf{p}_\mathrm{tot}|^3/24\pi^2$. 
Estimating the momentum volume $\phi\sim E^3$ with the energy scale $E$, we compute the timescale of this interaction to be $\mathcal{O}(1/G_FE^3)$. 
Intriguingly, this timescale is proportional to the estimate of Ref.~\cite{PhysRevD.110.123028} via the combination of a mean-field approximation $H\sim G_F\rho_\nu$ and a quasistatic equilibrium $\rho_\nu\sim E^3$; here, we have made neither of these approximations to arrive at the same estimate. 
In fact this result is quite general; when we consider more general initial conditions later, we find this interaction to have the same timescale. 

If we return to the question of scattering in the CoM frame, we find from $\mathbf{p}_\mathrm{tot}\to\mathbf{0}$ that $\phi(\mathbf{p}_\mathrm{tot})\to0$, and therefore, the evolution reduces to 
\begin{align}
    U_\mathrm{mom} \ket{\mathbf{p}_0,-\mathbf{p}_0} &\to 
    \left(\mathbbm{1} -itH_\mathrm{mom}\right)\ket{\mathbf{p}_0,-\mathbf{p}_0} 
    \nonumber \\
    &= \ket{\mathbf{p}_0,-\mathbf{p}_0} -it\sqrt{2}G_F\int \ket{\mathbf{p},-\mathbf{p}}\frac{\mathrm{d}\mathbf{p}}{(2\pi)^3}. 
\end{align}
The above result may appear to be nonunitary, but carefully note that our renormalized Hamiltonian becomes nilpotent in the CoM, where $\phi\to0$ in Eq.~\eqref{eq:hmom-pow-state}. 
A more intuitive and complete analysis of this problem 
can be broached if we 
treat our problem more generally with wave packets as opposed to only plane waves, as we proceed 
in the following section. 
{Nevertheless, we have seen that a calculation starting from the continuum interaction agrees with and complements both the numerical and analytical treatments on the lattice from our previous section. In particular, we have reiterated how approximate plane wave neutrino states interact on a timescale $1/\sqrt{2}G_F\phi$, although questions about the strength of their interaction would entail further consideration of wave packet size. } 

\section{Interacting Neutrino Wave Packets}
\label{sec:wavepackets}

A general formalism for calculating the oscillations of neutrinos modeled as plane waves or more generic wave packets was introduced by Kiers, Nussinov, and Weiss in Ref.~\cite{PhysRevD.53.537}. 
In their case, oscillations in vacuum and dense matter were considered. 
We adopt this approach and now consider the case of neutrinos interacting with each other. 
In this manner, we hope to gain a better understanding of how neutrino interactions manifest as a function of wave packet size and ultimately could survive from becoming trivial as per the earlier lattice calculation. 

Recall that a spherically symmetric Gaussian\footnote{Our analysis proceeds with this particular shape of wave function. However, we claim that our results are similar for more general shapes of wave functions that are still mostly confined to the same length scale as the Gaussian wave packet size. 
However, there may be important further considerations in an analysis of relativistic scattering of asymmetric wave packets, as argued for the case of hadrons in Ref.~\cite{Miller2025}. 
} wave packet may be written as 
\begin{align}
    \ket{\psi_\mathbf{p}} 
    &= \int \psi_\mathbf{p}(\mathbf{p^\prime})\ket{\mathbf{p^\prime}} \:\frac{\mathrm{d}\mathbf{p^\prime}}{(2\pi)^d} 
    \nonumber \\
    &= \frac{(2\pi)^{d/2}}{(\sqrt{2\pi}\sigma_p)^{d/2}} \int e^{-(\mathbf{p}-\mathbf{p^\prime})^2/4\sigma_p^2}\ket{\mathbf{p^\prime}} \: \frac{\mathrm{d}\mathbf{p^\prime}}{(2\pi)^d}
\end{align}
in $d$ spatial dimensions. 
Indeed, this description also includes regularized plane waves with quantization volume $V = \sigma_x^d\sim\sigma_p^{-d}$. For small wave packet sizes $\sigma_p\ll|\mathbf{p}|$, we find that the Gaussian wave packet behaves similarly to a Dirac distribution, $\exp[{-(\mathbf{p}-\mathbf{p^\prime})^2/4\sigma_p^2}] \approx (\sqrt{4\pi}\sigma_p)^d\delta(\mathbf{p}-\mathbf{p^\prime})$, representing the plane wave state $(\sqrt{2/\pi}\sigma_p)^{d/2}\ket{\mathbf{p}}$. 
Crucially, this basis is overcomplete, in the sense that
\begin{align}
    \braket{\psi_\mathbf{p}|\psi_\mathbf{p^\prime}} = e^{-(\mathbf{p}-\mathbf{p^\prime})^2/8\sigma_p^2}. 
\end{align}
In this way, the finite wave packet size 
sidesteps the need to regularize 
inner products of multiparticle states and thereby simplifies our analysis. 

We can then consider the evolution of interacting wave packets in this formalism with the Hamiltonian of Eq.~\eqref{eq:hmom}, 
\begin{align}
    H&_\mathrm{mom} \ket{\psi_{\mathbf{p}_0}, \psi_{\mathbf{q}_0}} = \frac{G_F}{\sqrt{2}} \int \psi_{\mathbf{p}_0}(\mathbf{p})\psi_{\mathbf{q}_0}(\mathbf{q}) 
    \nonumber \\
    &\times f(\mathbf{p},\mathbf{q}) f(\mathbf{p^\prime},\mathbf{p}+\mathbf{q}-\mathbf{p^\prime})^*  \ket{\mathbf{p},\mathbf{q}} \: \frac{\mathrm{d}\mathbf{p^\prime}}{(2\pi)^d} \, \frac{\mathrm{d}\mathbf{p}}{(2\pi)^d} \, \frac{\mathrm{d}\mathbf{q}}{(2\pi)^d}. 
\end{align}
The result is mostly the same as in Eq.~\eqref{eq:hmom-state}, though we now convolve that state with the wave packet shape of the incoming neutrinos $\psi_{\mathbf{p}_0}\psi_{\mathbf{q}_0}$. 
Further, the analogue to our idempotence property from the plane wave treatment is now 
\begin{align}
    H_\mathrm{mom}^n &\ket{\psi_{\mathbf{p}_0}, \psi_{\mathbf{q}_0}} 
    = \int \psi_{\mathbf{p}_0}(\mathbf{p})\psi_{\mathbf{q}_0}(\mathbf{q})  
    \nonumber \\
    &\times
    \left[\sqrt{2}G_F\phi(\mathbf{p}+\mathbf{q})\right]^{n-1}H_\mathrm{mom}\ket{\mathbf{p},\mathbf{q}} \: \frac{\mathrm{d}\mathbf{p}}{(2\pi)^d} \, \frac{\mathrm{d}\mathbf{q}}{(2\pi)^d},
\end{align}
implying for the time-evolved 
survival probability $P_s=|\!\braket{e^{-itH}}\!|^2$, 
in particular, 
\begin{align}
    &\braket{e^{-itH_\mathrm{mom}}}-1 
    =\nonumber \\
    &\int \psi_{\mathbf{p}_0}(\mathbf{p}) \psi_{\mathbf{p}_0}(\mathbf{p^\prime})^* \psi_{\mathbf{q}_0}(\mathbf{q}) \psi_{\mathbf{q}_0}(\mathbf{p}+\mathbf{q}-\mathbf{p^\prime})^* 
    \nonumber \\
    &\times \left(e^{-it\sqrt{2}G_F\phi(\mathbf{p}+\mathbf{q})}-1\right) \frac{f(\mathbf{p},\mathbf{q}) f(\mathbf{p^\prime},\mathbf{p}+\mathbf{q}-\mathbf{p^\prime})^*}{2\phi(\mathbf{p}+\mathbf{q})} \: 
    \nonumber \\
    &\times
    \frac{\mathrm{d}\mathbf{p^\prime}}{(2\pi)^d} \, \frac{\mathrm{d}\mathbf{p}}{(2\pi)^d} \, \frac{\mathrm{d}\mathbf{q}}{(2\pi)^d}. 
    \label{eq:gen-wp-scatter}
\end{align}

We can further analyze this form in the limits of small wave packet sizes $\sigma_p\gg|\mathbf{p}_0|,|\mathbf{q}_0|$ and large wave packet sizes $\sigma_p\ll|\mathbf{p}_0|,|\mathbf{q}_0|$. 
In the latter case, we recover a treatment of plane waves of quantization volume $V\sim\sigma_x^3\sim \sigma_p^{-3}$. 
Functionally, we find that the wave packets in momentum space are then well approximated by Dirac distributions, 
\begin{align}
    &\psi_{\mathbf{p}_0}(\mathbf{p}^\prime) \psi_{\mathbf{p}_0}(\mathbf{p})^* 
    \approx (2\pi)^{d/2} (4\pi\sigma_p)^d\, \delta(\mathbf{p}_0-\mathbf{p}) \delta(\mathbf{p}-\mathbf{p^\prime}),
\end{align}
and so our survival simplifies as 
\begin{align}
    \braket{e^{-itH_\mathrm{mom}}}-1 
    &\approx \left(4\pi\right)^d \left(\sqrt{2\pi}\sigma_p\right)^d 
    \nonumber \\
    &\times 
    \left(e^{-it\sqrt{2}G_F\phi(\mathbf{p}_0+\mathbf{q}_0)}-1\right) \frac{|f(\mathbf{p}_0,\mathbf{q}_0)|^2}{2\phi(\mathbf{p}_0+\mathbf{q}_0)}. 
\end{align}
Note that had we chosen general outgoing momenta for the wave packets $\mathbf{p}_0^\prime, \mathbf{q}_0^\prime$ with comparable sizes, we would still find amplitudes to approximate Dirac distributions 
and 
find a similar probability for arbitrary outgoing momenta. 
Indeed, in this case, as we take $\sigma_p\ll|\mathbf{p}|$, we can use Eqs.~\eqref{eq:hmom-state} and \eqref{eq:cont-evo} to estimate the (normalized) matrix elements for $\mathbf{p}_0^\prime+\mathbf{q}_0^\prime = \mathbf{p}_0+\mathbf{q}_0$, 
\begin{align}
    &\sigma_p^{2d}\braket{\mathbf{p}_0^\prime,\mathbf{q}_0^\prime|(U-\mathbbm{1})|\mathbf{p}_0,\mathbf{q}_0} \nonumber \\
    &= (e^{-it\sqrt{2}G_F\phi}-1)\frac{f(\mathbf{p}_0,\mathbf{q}_0)f(\mathbf{p}_0^\prime,\mathbf{q}_0^\prime)^*}{2\phi}\sigma_p^d. 
\end{align}

In order to estimate a quantity like the scattering amplitude from our above forms, we must also approximate the length in time of the interaction $t\sim \sigma_x\sim\sigma_p^{-1}$. 
If the box size is within the range $G_FE^2\ll \sigma_p/E \ll 1$,\footnote{Observe the width of this range for $E\sim\mathrm{MeV}$ spans 11 orders of magnitude. Let us also briefly remark that from observed supernova neutrino luminosities $L_\nu\sim10^{53}\:\mathrm{ergs}/\mathrm{s}$ we may roughly estimate with a bulb model~\cite{2010_DuanFullerQian}, $\rho_\nu\approx \sqrt{(L_\nu/E)/4\pi R_\nu^2}\sim \sigma_p^3$, that the limit $\sigma_p\lesssim E$ begins around $R_\nu\gtrsim100\:\mathrm{km}$. } then we find the above amplitude to reduce to $\sim G_F\sigma_p^2$. 
We arrive at the same result if we take the CoM limit ($\phi\to0$) without assuming this lower bound on the size of $\sigma_p$. 
Our cross section for neutrino wave packets $\Sigma_{\nu\nu}\sim G_F^2\sigma_p^4/E^2 = \Sigma_W(\sigma_p/E)^4$ appears similar to the well-known weak cross section $\Sigma_W := G_F^2E^2$. 

If we are to instead consider spatially narrow wave packets $\sigma_p\gg E$, then we may instead integrate over the large region in which incoming and outgoing Gaussian momentum distributions evaluated in $\braket{e^{-itH}}$ are overlapping and approximately constant $\sim \sigma_p^{-d/2}$. 
In this case, we may again linearly approximate our exponential since $t G_F E^3\ll 1$. 
Meanwhile, integrals over momenta each yield factors of $\sigma_p^d$. 
Thus, we again estimate $\braket{e^{-itH}}-1\sim G_F \sigma_p^2$, although here we see nonforward scattering to play a more significant role. 

It is tempting then to conclude that our result for the neutrino wave packet scattering {amplitude $\sim G_F\sigma_p^2$} is quite general, observing it for both narrow wave packets and plane waves. 
However, direct calculation of the above integrals for Eq.~\eqref{eq:gen-wp-scatter} in the regime $\sigma_p\sim E$ remains a more complicated task. 
Let us instead argue intuitively that in the limit of $\sigma_p\sim E$ we have taken our box size to be around the Compton wavelength. 
Precisely in this regime is where typical particle physics scattering arguments are already 
well established, and we would expect the usual weak cross section $\Sigma_{\nu\nu}\to\Sigma_W$ from our calculation $\sigma_p\to E$. 

{
Adopting the framework of treating the evolution of general wave packet sizes, we have been able to connect the behavior of interacting ``plane waves'' (functionally, $\sigma_p\ll E$) with that of ``wave packets'' ($\sigma_p\gg E$). 
In general, we find the timescale of the interaction to be dependent on the $\sqrt{2}G_F\phi$, where $\phi\sim E^3$ is the volume of momenta accessible in this interaction. 
In addition, we have obtained a sort of invariant formula for the amplitude of their interaction $\sim G_F\sigma_p^2$. 
}



\section{Ultrarelativistic Particles and Weak Interactions}
\label{sec:K+V}

Before we conclude, let us return to a discussion of a more complete Hamiltonian of interacting neutrinos, including not only the potential itself but also the kinetic energy, usually given in the form 
\begin{equation}
    H_\nu = \sum_{i=1}^{N_f}\int n_i(\mathbf{p}) E_i(\mathbf{p}) \:\frac{\mathrm{d}\mathbf{p}}{(2\pi)^d}, 
\end{equation}
where $n_i(\mathbf{p})=a_i^\dagger(\mathbf{p})a_i(\mathbf{p})$ is again the number occupancy operator for a neutrino of mass $m_i$, and $E_i(\mathbf{p})=\sqrt{\mathbf{p}^2+m_i^2}\approx|\mathbf{p}|[1+\mathcal{O}(m_i^2/|\mathbf{p}|^2)]$ is the relativistic energy of a free neutrino. 
Thus far, our analysis has been limited to the effects on neutrino wave packets of the interaction potential. 
Let us consider now a slightly more complete analysis with the Hamiltonian $H_\nu+H_{\nu\nu}$. 

First and most crucially, note the large separation of scales between these terms, $\lVert H_\nu\rVert\sim \mathrm{MeV}$ per a pair of neutrinos, whereas $\lVert H_{\nu\nu} \rVert \sim G_F \lVert H_\nu\rVert^3\sim 10^{-10}\:\mathrm{MeV}$ for a comparable pair. 
As such, we can expect that some sort of perturbative approaches should apply for wave packets as well as plane waves, and we consider both cases separately now. 
In short, we hope to indicate with this extra consideration that our earlier results focusing on $H=H_{\nu\nu}$ still provide valid physical insights to ultrarelativistic, interacting neutrinos. 

As we have seen in the previous section, the timescale on which the neutrino interaction varies is related to the wave packet size; $\tau\sim\sigma_x\sim\sigma_p^{-1}$. 
In the case of neutrino plane waves ($\sigma_p\ll E$), where the wave packet is much larger than the Compton wavelength of the particle, the effective potential should change slowly, thereby producing an adiabatic transition as neutrinos interact. That is, for plane waves, the evolution of the interaction satisfies the condition for adiabaticity, 
\begin{align}
\sigma_p &\sim \left\lvert \frac{1}{E_m-E_n} \Braket{E_m,t|\frac{\partial H}{\partial t}|E_n,t} \right\rvert 
\nonumber \\
&\ll \left\lvert \Braket{E_m,t|\frac{\partial}{\partial t}|E_n,t} \right\rvert \sim E, 
\end{align}
for energy levels $E_{m},E_n$ evolving in time. 
Consequently, if we take plane waves as our initial state, then we find our evolved wave function to remain an approximately definite kinetic energy eigenstate throughout the interaction and afterward ($t\gg\sigma_p^{-1}$), up to order $\mathcal{O}(G_F E^2)$. 
{Any error produced in results from this approximation of adiabaticity 
must be limited in size to $\mathcal{O}(\sigma_p/E)$ per our analysis above.} 

In contrast, narrow spatial wave packets ($\sigma_p\gg E$) of neutrinos would interact very much nonadiabatically, i.e., $\tau E \sim E/\sigma_p \ll 1$. 
Conveniently, in this case, we may simply perform a Suzuki-Trotter expansion of our time evolution operator, effectively separating our analysis of the potential from the kinetic term via 
\begin{align}
    e^{-i\tau (H_\nu+H_{\nu\nu})} = e^{-i\tau H_\nu} e^{-i\tau H_{\nu\nu}} + \mathcal{O}\left((\tau E)^2\right). 
\end{align}
We could thus isolate altogether the evolution operator of the interaction $\exp(-itH_{\nu\nu})$ considered in earlier sections {incurring an error in our short time evolution of $\mathcal{O}(E^2/\sigma_p^2)$ in this case}.

\section{Conclusions}
\label{sec:concl}


One might ask how the triviality of the continuum limit shown in our lattice problem can be resolved with well-known results for the weak interaction insofar as they may apply to neutrinos. 
To this question we must answer that we find the neutrino wave packet size to play a crucial role in the neutrino-neutrino interaction, and it therefore would be most appropriate to maintain a finite discretization in momentum space ${A}_p\sim\sigma_p\sim V^{-1/3}$ for the lattice model or otherwise to appropriately sum over the growing number of lattice sites within a wave packet's width as the lattice grows finer. 
In our CoM problem presented earlier, this fixture would amount to selecting a finite number of momentum pairs $M\sim\rho_\nu/E^3$, in which case oscillations from neutrino-neutrino interactions are still nonvanishing. 

Additionally, in the limit of interacting plane waves, we have shown that the forward scattering (i.e., flavor swapping) will be 
most physically interesting, as the invariant scattering amplitude (or, equivalently, cross section) is directly related to the spread in momentum space, which is small by assumption here, and therefore may have little bearing on kinematics. 
Conversely, in the limit of small neutrino wave packets, we see that nonforward scattering plays a larger role. 
The analysis presented above considers a picture of two ultrarelativistic neutrinos scattering via the weak interaction, using varied wave packet sizes. 
It is therefore somewhat generic in helping to better understand neutrino-neutrino interactions, but it is also simplistic and leaves room for further analysis along various directions. 

While we have considered the effects of including more allowed momentum states in our simulations of the neutrino interaction, we have not yet comprehensively probed the effects of including more neutrinos occupying those states in simulations simultaneously. 
As such, the consequences of this analysis through the lens of a many-body theory still need to be seen in future work. 
{The evolution of more than two interacting neutrino wave packets could be much more complicated to predict, yet it could induce more intricate or emergent behaviors yet to be uncovered; without further study, we cannot yet say what consequences the inclusion of wave packet nature would have on many-body neutrino physics. 
} 
Nevertheless, we maintain that an analysis with two neutrinos can say much about the physics of this interaction. 
The interaction of neutrinos with antineutrinos is not directly addressed in our analysis and opens up greater complexity 
to be investigated, as would an explicit treatment of the spin nature of neutrinos and antineutrinos alike. 
Although the generalization of our analysis to arbitrary numbers of neutrino flavors is straightforward since we did not depend upon a choice of $N_f$ in this work, there may be more features of entanglement between momenta and flavor yet to be appreciated.

\section*{Acknowledgments}

We thank Yukari Yamauchi for assistance in adapting an earlier program from Github for use in numerical simulations performed in this work. 
We thank Wouter Dekens and Yukari Yamauchi also for their verification of several analytic results presented in this work. 
We thank Baha Balantekin, Aurel Bulgac, Joseph Carlson, Andrea Carosso, Vincenzo Cirigliano, Wouter Dekens, Julien Froustey, Lucas Johns, Joshua Martin, Gerald A.~Miller, Sherwood Richers, and Yukari Yamauchi for valuable discussions while 
this work was done. 
M.J.C.~is supported by the U.S.~Department of Energy Grant No.~DE-FG02-97ER-41014 (U.W.~Nuclear Theory).

\section*{Data Availability} 

The data that support the findings of this article are not publicly available upon publication because it is not technically feasible and/or the cost of preparing, depositing, and hosting the data would be prohibitive within the terms of this research project. The data are available from the author upon reasonable request. 

\appendix

\section{Volume of Outgoing Momentum Space}
\label{app:vol-mom}

In Sec.~\ref{sec:lattice}, we introduce the space of outgoing momentum states permitted by the weak interaction of neutrinos with incoming momenta $\mathbf{p}_0+\mathbf{q}_0=\mathbf{p}_\mathrm{tot}$. 
Here, we show the calculation of volume for this space, $\phi$, defined in Eq.~\eqref{eq:mom-vol}. 

We begin by recognizing the dependence of this volume on the geometry of incoming momenta, 
\begin{align}
    \phi(\mathbf{p}_\mathrm{tot}) 
    &= 
    \frac{1}{2}\int \left(1-\cos\theta_{\mathbf{p^\prime},\mathbf{p}_\mathrm{tot}-\mathbf{p^\prime}}\right)
    \:\frac{\mathrm{d}\mathbf{p^\prime}}{(2\pi)^d} 
\end{align}
while 
\begin{align}
    \cos\theta_{\mathbf{p^\prime},\mathbf{p}_\mathrm{tot}-\mathbf{p^\prime}} 
    &= \frac{\mathbf{p^\prime}\cdot(\mathbf{p}_\mathrm{tot}-\mathbf{p^\prime})}{|\mathbf{p^\prime}| \, |\mathbf{p}_\mathrm{tot}-\mathbf{p^\prime}|} 
    \nonumber \\
    &= \frac{|\mathbf{p}_\mathrm{tot}|\cos\theta_{\mathbf{p^\prime},\mathbf{p}_\mathrm{tot}}-|\mathbf{p^\prime}|}{|\mathbf{p}_\mathrm{tot}-\mathbf{p^\prime}|},
\end{align}
so we have the simple formula, 
\begin{align}
    \phi(\mathbf{p}_\mathrm{tot}) 
    = \frac{1}{2}\int& \left(1 - |\mathbf{p}_\mathrm{tot}|\frac{\cos\theta_{\mathbf{p^\prime},\mathbf{p}_\mathrm{tot}}}{|\mathbf{p}_\mathrm{tot}-\mathbf{p^\prime}|} + \frac{|\mathbf{p^\prime}|}{|\mathbf{p}_\mathrm{tot}-\mathbf{p^\prime}|}\right)
    \nonumber \\
    &\times 
    \frac{\mathrm{d}\mathbf{p^\prime}}{(2\pi)^d}. 
    \label{eq:mom-vol-geoints}
\end{align}
Now, choosing to renormalize this integral using a high-momentum cutoff $|\mathbf{p^\prime}|\leq\Lambda$, taken such that $\Lambda>|\mathbf{p}_\mathrm{tot}|$ as well, we can evaluate the integrals of each of the three integrand terms in Eq.~\eqref{eq:mom-vol-geoints} with spherical geometry, 
\begin{align}
    \int_{\{|\mathbf{p^\prime}|\leq\Lambda\}} \mathrm{d}\mathbf{p^\prime} &= \frac{4\pi}{3}\Lambda^3, 
    \\
    \int_{\{|\mathbf{p^\prime}|\leq\Lambda\}} \frac{\cos\theta_{\mathbf{p^\prime},\mathbf{p}_\mathrm{tot}}}{|\mathbf{p}_\mathrm{tot}-\mathbf{p^\prime}|} 
    \: \mathrm{d}\mathbf{p^\prime} &= \frac{4\pi}{3}|\mathbf{p}_\mathrm{tot}|\Lambda-\pi|\mathbf{p}_\mathrm{tot}|^2,  
    \\
    \int_{\{|\mathbf{p^\prime}|\leq\Lambda\}} \frac{|\mathbf{p^\prime}|}{|\mathbf{p}_\mathrm{tot}-\mathbf{p^\prime}|} 
    \: \mathrm{d}\mathbf{p^\prime} &= \frac{4\pi}{3}\Lambda^3 -\frac{\pi}{3}|\mathbf{p}_\mathrm{tot}|^3. 
\end{align} 
Explicitly, the latter terms can be calculated naturally in a piecewise fashion over the complementary regions $\{|\mathbf{p^\prime}|\leq|\mathbf{p}_\mathrm{tot}|\}$ and $\{|\mathbf{p}_\mathrm{tot}|<|\mathbf{p^\prime}|<\Lambda\}$, as 
\begin{align}
    \int& \frac{|\mathbf{p^\prime}|}{|\mathbf{p}_\mathrm{tot}-\mathbf{p^\prime}|} 
    \: \mathrm{d}\mathbf{p^\prime} 
    \nonumber \\
    &= \frac{2\pi}{|\mathbf{p}_\mathrm{tot}|} \int \big( |\mathbf{p^\prime}|+|\mathbf{p}_\mathrm{tot}| 
    -\left||\mathbf{p^\prime}|-|\mathbf{p}_\mathrm{tot}|\right| \big)
    \: |\mathbf{p^\prime}|^2\mathrm{d}|\mathbf{p^\prime}|,
\end{align}
\begin{align}
    &\int \frac{\cos\theta_{\mathbf{p^\prime},\mathbf{p}_\mathrm{tot}}}{|\mathbf{p}_\mathrm{tot}-\mathbf{p^\prime}|} 
    \: \mathrm{d}\mathbf{p^\prime} 
    \nonumber \\
    &= \frac{2\pi}{3|\mathbf{p}_\mathrm{tot}|^2} \int \bigg[ \left(|\mathbf{p^\prime}|^2-|\mathbf{p^\prime}||\mathbf{p}_\mathrm{tot}|+|\mathbf{p}_\mathrm{tot}|^2\right) \left(|\mathbf{p^\prime}|+|\mathbf{p}_\mathrm{tot}|\right) 
    \nonumber \\
    &\phantom{=}
    -\left(|\mathbf{p^\prime}|^2+|\mathbf{p^\prime}||\mathbf{p}_\mathrm{tot}|+|\mathbf{p}_\mathrm{tot}|^2\right) \left||\mathbf{p^\prime}|-|\mathbf{p}_\mathrm{tot}|\right| \bigg]
    \: \mathrm{d}|\mathbf{p^\prime}|.
\end{align}
In summary, our momentum volume with a cutoff $\Lambda$ becomes 
\begin{align}
    \phi(\mathbf{p}_\mathrm{tot}) = \frac{1}{2(2\pi)^3} \left( \frac{8\pi}{3}\Lambda^3 - \frac{4\pi}{3}\Lambda|\mathbf{p}_\mathrm{tot}|^2 + \frac{2\pi}{3}|\mathbf{p}_\mathrm{tot}|^3\right). 
\end{align}
Intuitively, our final result must be independent of the scheme of renormalization. 
{More precisely, we should expect that the effective field theory of the Fermi four-point interaction with a hard momentum cutoff also involves a cutoff dependence in our Fermi coupling $G_F(\Lambda)$. Ultimately, one can show that, in allowing for this dependence, we also introduce counterterms to our Hamiltonian that are dependent on the cutoff $\Lambda$ and include higher order derivatives producing $\mathcal{O}(G_F|\mathbf{p}|^2)$ corrections, which we would demand to cancel any such $\Lambda$-dependence in the calculation of relevant observables. 
} 
Thus, we can 
{identify} 
{a renormalized value for the volume of our allowed momentum space with the cutoff-independent contribution,} 
\begin{align}
    \phi(\mathbf{p}_\mathrm{tot}) \to \frac{|\mathbf{p}_\mathrm{tot}|^3}{24\pi^2}. 
\end{align}

\bibliography{nunu}

\end{document}